\documentclass[manuscript]{acmart}
\AtBeginDocument{%
  \providecommand\BibTeX{{%
    \normalfont B\kern-0.5em{\scshape i\kern-0.25em b}\kern-0.8em\TeX}}}

\setcopyright{acmlicensed}
\copyrightyear{2024}
\acmYear{2024}
\acmDOI{XXXXXXX.XXXXXXX}

\acmConference[DIS '24]{Make sure to enter the correct conference title from your rights confirmation email}{July 01--05, 2024}{Copenhagen, Denmark}
\begin{document}

\title{Using Tangible Interaction to Design Musicking Artifacts for Non-musicians}

\author{Luc\'ia Montesinos}
\affiliation{%
  \institution{IT University of Copenhagen}
  \streetaddress{Rued Langgaards Vej 7}
  \city{Copenhagen}
  \country{Denmark}}
\email{lumo@itu.dk}
\orcid{0009-0002-1438-6461}

\author{Halfdan Hauch Jensen}
\affiliation{%
  \institution{IT University of Copenhagen}
  \streetaddress{Rued Langgaards Vej 7}
  \city{Copenhagen}
  \country{Denmark}}
\email{halj@itu.dk}
\orcid{0000-0002-0351-6489}

\author{Anders Sundnes L\o vlie}
\affiliation{%
  \institution{IT University of Copenhagen}
  \streetaddress{Rued Langgaards Vej 7}
  \city{Copenhagen}
  \country{Denmark}}
\email{asun@itu.dk}
\orcid{0000-0003-0484-4668}

\renewcommand{\shortauthors}{}

\begin{abstract}

This paper presents a Research through Design exploration of the potential for using tangible interactions to enable active music experiences - \textit{musicking} - for non-musicians. We present the \textit{Tubularium} prototype, which aims to help non-musicians play music without requiring any initial skill. We present the initial design of the prototype and the features implemented in order to enable music-making by non-musicians, and offer some reflections based on observations of informal initial user explorations of the prototype.

\end{abstract}



\begin{teaserfigure}
  \includegraphics[width=\textwidth]{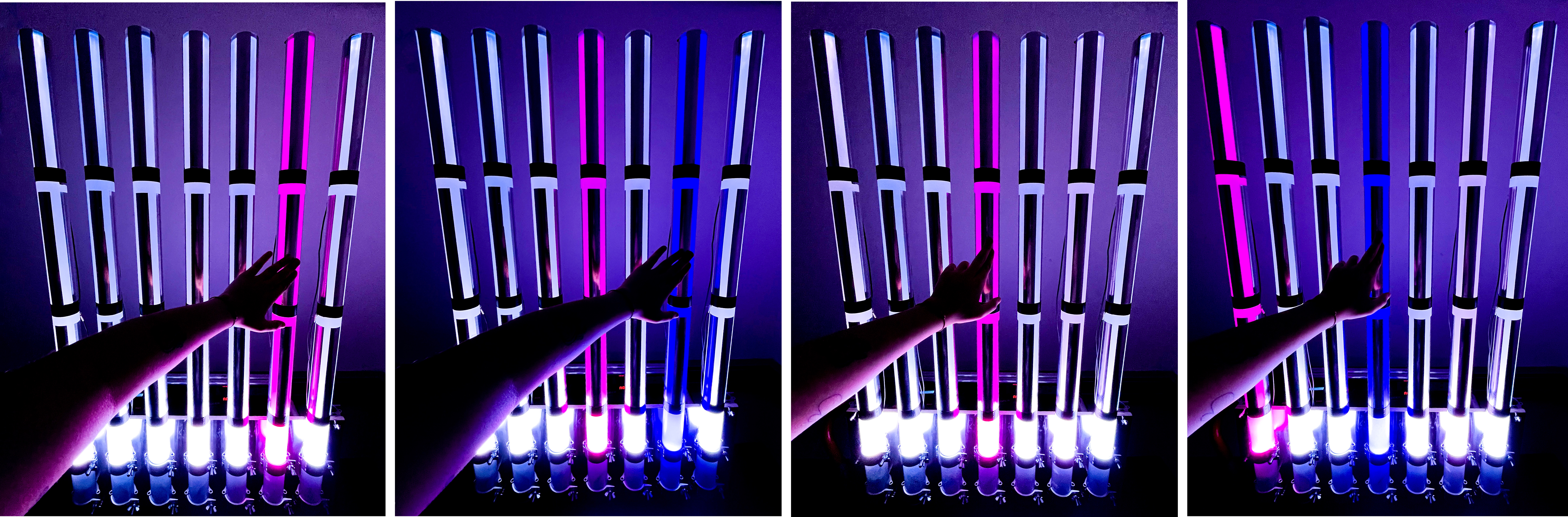}
  \caption{Chord progression mode in the Tubularium. The pink light indicates the next chord in the progression. When the tube is touched, the light turns blue and the system advances to the next chord, illuminating in pink the corresponding tube.}
  \label{fig:teaser}
\end{teaserfigure}


\maketitle
\section{Introduction}

While advances in music technology have brought rich opportunities for experiencing music and facilitated a so-called music democratization process, they have also brought a \textit{passivization} in how we experience music 
\cite{jensenius_sound_2022}, treating music more as a medium for passive listening than as an \textit{activity} to participate in.
In this study, we draw inspiration from Christopher Small's argument that music should be an active process for anyone involved in the experience: \textit{musicking} \cite{small_musicking_1998}. For Small the performance is central to the experience of music: "a performance does not exist in order to present musical works, but rather, musical works exist in order to give performers something to perform" \cite[p.8]{small_musicking_1998}. Those who play instruments or compose have inherent access to these active experiences. However, mastering a traditional instrument may require formal training and many years of practice, posing a large obstacle to active participation in music making \cite{merriam_learning_2007}. 

We present an ongoing Research through Design project exploring how to create new musical interfaces that open the opportunity for non-musicians to perform music by mediating the process of music-making. Here, the Tubularium is introduced as a musicking artifact: a tangible interface with features tailored to support active music-making. 

\section{Related work}


Research in Human-Computer Interaction (HCI) has long been exploring the design of New Interfaces for Musical Expression (NIME) \cite{Fasciani_2021}, often through embodied and tangible interactions \cite{mice_super_2022, tapparo_2022, Cotton_2021, reactable2017}. Designing tangible interfaces involves designing for the digital and physical as well as for the relationships between these spaces. There are multiple studies on tangible interactions for music interfaces such as tabletops \cite{jorda_reactable_2007, franceschini_laney_2021, patten_audiopad_2002} and blocks \cite{Schiettecatte_audiocubes_2008, wu_musicking_2019}. 
According to Jordà \cite{jorda_digital_2005}, much of the research in NIME has tended to target expert users and focus on developing interfaces that permit improvisation, experimentation, and the performance of a range of different pieces of music. There has been however an emerging trend in NIME about exploring how to facilitate active experiences in music making for non-musicians. Murray-Browne \cite{murray-browne_interactive_2012} suggests a distinction between Digital Music Instruments (DMI) - denoting the hitherto dominant paradigm in NIME - and Interactive Music Systems (IMS), which are systems built for non-expert users and that take input from such users and respond with music. 
Murray-Browne highlights Gelinek and Serafin's work on explorability \cite{gelineck_practical_2010}, suggesting that an IMS should encourage explorative behavior whilst being intuitive enough to give the confidence to continue.

An early example of such a system was the Hotz Box \cite{HotzBox} released in the 1990s as a midi controller designed for both novices and experienced musicians. It featured a translator software that aimed to remove the need of knowing music theory. Although the Hotz Box itself didn't succeed as a product, the concept of music translation is now a common feature in current devices with grid controllers \cite{Ashun_2021, Squarp_2022, Novation_2021}. We can also find in the current market sequencers that aim to support explorative behaviour, but these turn out to be either overly complicated for beginners \cite{Conductive_2020} or meant to be used as a compositional tool \cite{Roland_2022}. A remarkable device is Dato Duo \cite{Dato_2017}, a sequencer and synthesizer designed for all ages. However, in their efforts to simplify music making, they limited the musical possibilities to a single pentatonic scale. In this study we will explore the possibility of developing a musicking artifact that is as simple to use for non-musicians as the Dato Duo, but which allows a much greater breadth of musical expression.

\section {Approach}
\label{sec:approach}

This project has followed a Research Through Design \cite{gaver_what_2012} approach in which we explore the insights that come out of the design process and the designed artifact. In the following we will first present the design rationale and process, and later on the features of the current prototype, followed by a discussion to draw out salient insights from the design.



The Tubularium was designed through the musicking lens and with the aim of facilitating music-making for non-musicians. The design of the artifact was driven by three main interaction principles:
\begin{itemize}
    \item the system should help non-musicians create melodious sounds without requiring any initial skill
    \item the interaction should elicit a sense of agency and ownership of the sounds generated
    \item the design should facilitate a meaningful music experience without aiming to educate the users
\end{itemize}
In the following we will present the design of the Tubularium prototype, and discuss the design in light of these three principles.  

\section{Tubularium}
\begin{figure}[!tbp]
  \centering
  \begin{minipage}[b]{0.55\textwidth}
    \includegraphics[width=\textwidth]{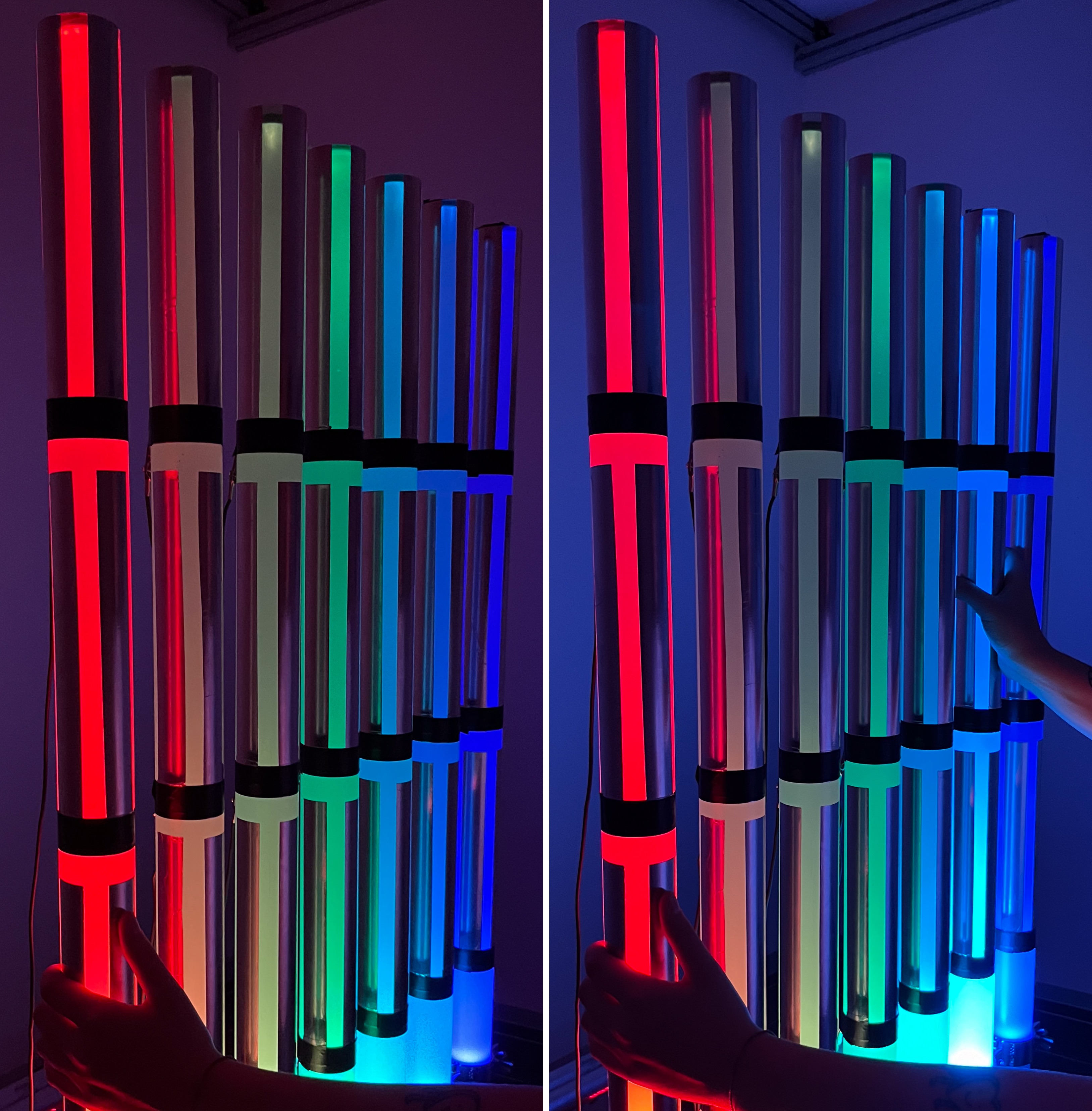}
    \caption{Single note mode in the Tubularium. When a tube is touched, the light inside shines brightly. Several tubes can be touched at a time.}
    \label{fig:instA_temp}
  \end{minipage}
  \hfill
  \begin{minipage}[b]{0.35\textwidth}
    \includegraphics[width=\textwidth]{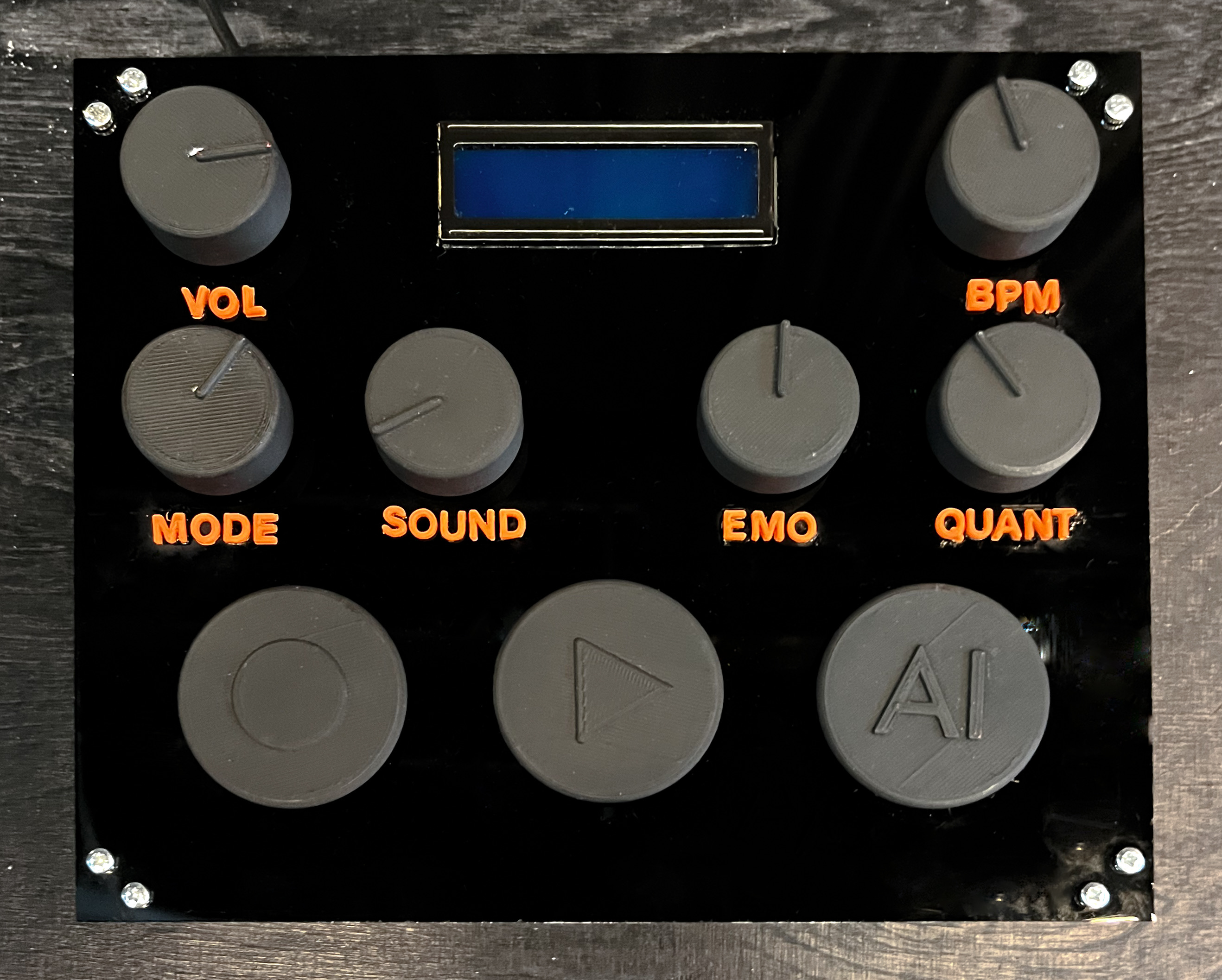}
    \caption{Controlling interface}
    \label{fig:cont_temp}
  \end{minipage}
\end{figure}

The Tubularium, shown in Figure \ref{fig:instA_temp}, is a melodic instrument consisting of 7 tubes and 21 sensing areas, as well as a controlling interface, shown in Figure \ref{fig:cont_temp}. 

The physical instrument is made out of seven acrylic opal tubes standing on an aluminium structure. On the surface of the tubes there are three distinct sensing areas, delimited by vertical stripes of electrical conductive tape. Each sensing area is connected to a channel of a capacitive touch sensor breakout, which is connected to an Arduino board. Each tube contains inside a RGB LED connected to a DMX controller. The controlling interface, made of acrylic laminates, contains a LCD display, potentiometers and push buttons which are connected to another Arduino board. 

The instrument is controlled by a custom software made using the visual programming language Max/MSP \cite{cycling_2023}. It comprises a main patch with three sub-patches: one for communication handling via serial ports with a custom protocol, and the others for single note and chord progression modes, managing the dynamic mapping and additional instrument features.

The Tubularium allows novices to easily play melodious sounds through a 'chord progression' mode, as shown in Figure \ref{fig:teaser}. When the user touches the sensing area the instrument plays a harmonious chord with the corresponding note as base, and the light within the tube lights up blue. Both the sound and the light will last for as long as the tube is being touched. In this mode, the lights inside the tubes are also used as indicators of the next chord in the progression: The next tube the player should touch shines pink, in order to help the players play chords in the right order. This mode supports the exploration of chord progressions rhythmically. It also invites the player to explore their embodiment regarding the relationship between the progressions and the emotions they evoke in them.

Furthermore, the instrument can also be played in single note mode, where each input triggers a single note. Several tubes can be touched at a time, like shown in \ref{fig:instA_temp}. This mode allows the user to explore melodies, create a lead or even improvise a solo, through real time play, history keeping mechanism, and co-creation with an AI.

\subsection{Instrument features}
\subsubsection{Dynamic mapping}
The 7 tubes in the Tubularium are representing each of the notes within a heptatonic scale. The tubes's three sensing areas correspond to octaves. The upper area is one octave higher than the middle one and two octaves higher than the bottom one. This facilitates the musicking process by removing the need for non-expert users to grasp these music theory notions in order to play only the notes within a specific scale. The mapping was done dynamically so that it adapts to the popular heptatonic scales starting by any root note. 
\subsubsection{Emotion exploration}
The emotion knob in the controller invites players to explore a range of emotions. It is widely acknowledged that scales and chord progressions can help evoke specific emotions. Although the relationship between these is complex and shaped by our embodiment, musicians often use general associations between emotions and scales to help evoke emotions. Turning the knob adjusts the mapping of the tubes to a new scale and root note, and a new chord progression is loaded. This feature encourages the user to explore how different chord progressions can evoke different emotions.
 \subsubsection{History-keeping}
The user can record and play back their musicking, and use this as a way of understanding the sounds made or to create more complex sounds. History keeping mechanisms have been reported as creativity support tools \cite{wu_design_2018}, as having the possibility of revisiting the created output makes creative improvements easier to achieve \cite{shneiderman_creativity_2007}. The history keeping mechanism for this instrument is set to only save the latest recording to prompt the user to focus on live playing rather than on sequencing or remixing. 
\subsubsection{AI jamming}
There is also the possibility of co-creating music with an AI. To do so, the user needs to record themselves and then press the AI button. The recording is then sent to a Hidden Markov Model based on the MAX/MSP object [ml.markov] \cite{Smith2012a}. The model returns an endless improvisation based on the recording data. This improvisation can be played back simultaneously as the user keeps playing, allowing them to jam with themselves to create richer melodies or to be able to play both instrument modes. This is also a feature that can facilitate music co-creation with others, as a user could be jamming with an AI trained on the previous user style.
\subsubsection{Quantization}
A quantization feature was also introduced to support the rhythm. The user can choose how much support they want by turning a knob, which in returns delays the triggering of the sounds to match the user's defined subdivision of the beat.
\subsubsection{Sound exploration}
Finally the user is also able to experiment with the sounds the instrument makes. The user can choose which sounds the instrument makes through the controller. This encourages experimentation fuelling the creative process.

\section{Discussion}
 


Here we reflect on the design of Tubularium in light of the three design principles set out above (see Section \ref{sec:approach}). As this is a work in progress we have not yet carried out a rigorous evaluation of the design. However, the design has been presented to colleagues and students in internal events in our research group as well as various informal meetings, and the reflections here are based on observations of how people engage with the prototype in these contexts. The participants in these informal exchanges are typically researchers and students working with design, HCI and computer science. While our group includes some who are active musicians, for these reflections we focus on the interactions of non-musicians with the prototype. 

\textit{1. The system should help non-musicians create melodious sounds without requiring any initial skill.}

In a recent lab event we observed numerous students and outside guests go up to the Tubularium and start exploring its sounds on their own. The physical setup with the large tubes and the lights seem to invite users to explore and play. While users with knowledge of music may start to play melodies based on their memory and their understanding of the scales, non-musician users may end up playing notes in more random sequences and octaves. However, since the notes are restricted to the heptatonic scales, even the sounds made by non-musicians are harmonious. Furthermore, non-musicians can use the chord progression indicators to play melodious sequences of chords. As the relationship between the user and the instrument evolves, more complex sounds can be made.

\textit{2. The interaction should elicit a sense of agency and ownership of the sounds generated.}

This is a common challenge when designing digital music instruments, as there is an inherent separation between the action-sound pairing. In the case of the Tubularium, the focus in this regard was set on ensuring that the user is aware that they have choices - and that their actions have an effect over the sounds produced. In the observations made during the lab event, users would put this principle to the test by gliding their hand across the tubes like a harp, or by trying to touch all of the tubes at the same time. Both of these actions are supported by the prototype. However, we also observed how some of them tried to change the dynamics of the sound by changing the intensity of the touch, something which is not taken into account by the prototype in its current state. Thus, some users may have experienced a misalignment between their action and the sound produced.

\textit{3. The design should facilitate a meaningful music experience without aiming to educate the users.}

Whether Tubularium produces meaningful music experiences or not is not easy to determine based on the short, informal observations we have reflected on here. This question can only be answered through a more thorough evaluation. In this evaluation we will aim to explore whether the Tubularium is indeed a musicking artifact; in other words, whether it provides an active and embodied encounter with music. Furthermore, the evaluation will also aim to provide insight about whether the ideas or feelings elicited by the artifact are aligned with the other two principles it was designed after. Future work will also include an evaluation focused on the different features present in the Tubularium to evaluate which of these features contribute to facilitating music-making for non-musicians.





\bibliographystyle{ACM-Reference-Format}
\bibliography{main}

\end{document}